

%
\input epsf
%
%
\input harvmac
\noblackbox
\def\Title#1#2{\rightline{#1}\ifx\answ\bigans\nopagenumbers\pageno0\vskip1in
\else\pageno1\vskip.8in\fi \centerline{\titlefont #2}\vskip .5in}

scaled\magstep3
 
scaled\magstep3
%
%
\ifx\epsfbox\UnDeFiNeD\message{(NO epsf.tex, FIGURES WILL BE IGNORED)}
\def\figin#1{\vskip2in}
\else\message{(FIGURES WILL BE
INCLUDED)}\def\figin#1{#1}
\fi
\def\Fig#1{Fig.~\the\figno\xdef#1{Fig.~\the\figno}\global\advance\figno
 by1}
%
%
%
%
\def\ifig#1#2#3#4{
\goodbreak\midinsert
\figin{\centerline{\epsfysize=#4truein\epsfbox{#3}}}
\narrower\narrower\noindent{\footnotefont
{\bf #1:}  #2\par}
\endinsert
}
%
%

\def\ajou#1&#2(#3){\ \sl#1\bf#2\rm(19#3)}

\def\frac#1#2{{#1 \over #2}}

\def\a{\alpha}
\def\p{\partial}
\def\r{{\tilde r}}

\def\to{{\rightarrow}}

\def\B{{\widehat B}}
\def\vp{{\varphi}}
\def\tp{{\tilde \varphi}}
\def\({\left (}
\def\){\right )}
\def\[{\left [}
\def\]{\right ]}
\def\KK{{Kaluza-Klein}}
\def\q{{\widehat q}}

%
%
\lref\dggh{H.F. Dowker, J.P. Gauntlett, S. B. Giddings and G. Horowitz,
\ajou Phys. Rev. &D50 (94) 2662.}
\lref\dgkt{H.F. Dowker, J.P. Gauntlett, D. A. Kastor, J. Traschen,
\ajou Phys. Rev. &D49 (94) 2909.}
\lref\andy{D. Garfinkle, S. Giddings and A. Strominger, ``Entropy in Black
Hole Pair Production'', Santa Barbara preprint UCSBTH-93-17,
gr-qc/9306023.}
\lref\gm{G.W. Gibbons and K. Maeda,
\ajou Nucl. Phys. &B298
(88) 741.}
\lref\ghs{D. Garfinkle, G. Horowitz, and A. Strominger,
\ajou Phys. Rev. &D43 (91) 3140, erratum\ajou Phys. Rev.
& D45 (92) 3888.}
\lref\gwg{G.W. Gibbons,
in {\sl Fields and geometry}, proceedings of
22nd Karpacz Winter School of Theoretical Physics: Fields and
Geometry, Karpacz, Poland, Feb 17 - Mar 1, 1986, ed. A. Jadczyk (World
Scientific, 1986).}
\lref\garstrom{D. Garfinkle and A. Strominger,
\ajou Phys. Lett. &256B (91) 146.}
\lref\ernst{F. J. Ernst, \ajou J. Math. Phys. &17 (76) 515.}
\lref\gh{G. W. Gibbons and S. W. Hawking, \ajou Commun. Math. Phys. &66 (79)
291.}
\lref\rafff{R. D. Sorkin, \ajou Phys. Rev. &D33 (86) 978.}
\lref\bomb{L. Bombelli, R. K. Koul, G. Kunstatter, J. Lee and R. D. Sorkin,
\ajou Nucl. Phys. &B289 (87) 735.}
\lref\rafKK{R. D. Sorkin, \ajou Phys. Rev. Lett. &51 (83) 87; {\bf 54} (1985)
86{\bf E}.}
\lref\grossperry{D. Gross and M.J. Perry, \ajou Nucl.Phys. &B226 (83) 29.}
\lref\schwinger{J. Schwinger}
\lref\melvin{M. A. Melvin, \ajou Phys. Lett. &8 (64) 65.}
\lref\myp{R. C. Myers and M. J. Perry,
\ajou Ann. Phys. &172 (86) 304.}
\lref\witten{E. Witten, \ajou Nucl. Phys. &B195 (82) 481.}
\lref\ginsperry{P. Ginsparg and M.J. Perry, \ajou Nucl.Phys. &B222 (83) 245.}
\lref\gott{J.R. Gott, \ajou Nuov. Cim. &22B (74) 49.}
\lref\schulman{L.S. Schulman, \ajou Nuov. Cim. &2B (71) 38.}
\lref\peres{A. Peres, \ajou Phys. Lett &31A (70) 361.}
\lref\banks{T. Banks, M. Dine, H. Dijkstra and W. Fischler, \ajou Phys. Lett
&212B (88) 45.}
\lref\sen{A. Sen, \ajou Int. J. Mod. Phys. &A9 (94) 3707.}
\lref\wit{E. Witten, \ajou Nucl. Phys. &B443 (95) 85.}
\lref\hullt{C. M. Hull and P. K. Townsend, \ajou Nucl. Phys. &B438 (95) 109.}
%
%
\Title{\vbox{\baselineskip12pt
\hbox{UCSBTH-95-15}
\hbox{CALT-68-2005}
\hbox{hep-th/9507143}}}
{\vbox{\centerline{The Decay of Magnetic Fields in Kaluza-Klein Theory
}}}
{
\baselineskip=12pt
\centerline{Fay Dowker$^{1a}$, Jerome P. Gauntlett$^{2}$,
Gary W. Gibbons$^{3}$ and Gary T. Horowitz$^{1b}$}
\bigskip
\centerline{\sl $^1$Department of Physics}
\centerline{\sl University of California}
\centerline{\sl Santa Barbara, CA 93106}
\centerline{\it $^a$Internet: dowker@cosmic.physics.ucsb.edu}
\centerline{\it $^b$Internet: gary@cosmic.physics.ucsb.edu}
\bigskip
\centerline{\sl $^2$California Institute of Technology}
\centerline{\sl Pasadena, CA, 91125}
\centerline{\it Internet: jerome@theory.caltech.edu}
\bigskip
\centerline{\sl $^3$DAMTP, University of Cambridge}
\centerline{\sl Silver St., Cambridge, CB3 9EW}
\centerline{\it Internet: gwg1@anger.amtp.cam.ac.uk.edu}
\medskip
\centerline{\bf Abstract}
Magnetic fields in five-dimensional Kaluza-Klein theory
compactified on a circle correspond to ``twisted'' identifications
of five dimensional Minkowski space.
We show that a  five dimensional generalisation of
the Kerr solution can be analytically continued to
construct an instanton that gives rise to two
possible decay modes of a magnetic field.
One decay mode is the generalisation of the ``bubble decay"
of the Kaluza-Klein vacuum described
by Witten. The other decay mode, rarer for weak fields,
corresponds in four dimensions to the
creation of monopole-anti-monopole pairs.
An instanton for the latter process is already known and
is given by the analytic
continuation of the \KK\ Ernst metric, which we show
is identical to
the five dimensional Kerr solution.
We use this fact to
illuminate further properties of the decay process.
It appears that fundamental fermions can eliminate
the bubble decay of the magnetic field, while allowing
the pair production of Kaluza-Klein monopoles.
}


\Date{6/95}
%

\newsec{Introduction}
The standard  Kaluza-Klein vacuum, $M^4 \times
S^1$, is known to be unstable. Witten showed \witten\
that it can semiclassically decay by nucleating a
``bubble of nothing'' which appears to expand into space.
In this paper we will consider
another class of ``vacua" in Kaluza-Klein theory which correspond to static
magnetic
flux tubes in four dimensions. Although these solutions are non-trivial
four dimensional configurations they are simply obtained from
dimensional reduction of
five dimensional Minkowski space, $M^5$,
with  ``twisted'' identifications. We will see that
these backgrounds are also unstable, and in fact, have two different decay
modes. The first was discussed in \dggh\ and corresponds to the pair
creation of Kaluza-Klein monopoles. We will show that there is another
decay mode which occurs at a much higher rate. It is
a direct generalization of the
``expanding bubble'' found
by Witten.

The semiclassical decay of a vacuum can be described by an instanton, i.e.,
a euclidean solution to the field equations which interpolates between the
initial and final states.  The leading approximation to the decay rate
is  simply $e^{-I}$ where $I$ is the instanton action.
To show that $M^4 \times S^1$ is unstable,
Witten constructed an appropriate instanton by analytically continuing
the five dimensional Schwarzschild solution. We will use the five dimensional
Kerr solution to construct an instanton which describes the decay of
a Kaluza-Klein magnetic field. The subsequent evolution can be obtained by
a further analytic continuation (as in the Schwarzschild case) and
resembles an expanding bubble.

In \dggh, an instanton describing the
pair creation of monopoles was constructed by analytically
continuing the Kaluza-Klein Ernst
solution \dgkt. Recall that in four dimensional Einstein-Maxwell
theory, the Ernst solution \ernst\ describes
two oppositely charged black holes accelerating
apart in a background magnetic field.
Since the Kaluza-Klein monopole
\refs{\rafKK,\grossperry} is just an extremal
magnetically charged black hole, the pair creation of monopoles
can be described using
an instanton constructed from the Kaluza-Klein analog of the Ernst solution.

Remarkably enough,  it turns out that the instanton we
construct from the Kerr metric is identical to the one previously constructed
from the Ernst solution! At first sight, this appears impossible. Not only
does the spacetime containing an expanding bubble seem very different
from one containing two accelerating monopoles, but
the actions for the two instantons are different: in the limit  that
the asymptotic magnetic field $B \to 0$,
the rate for monopole creation vanishes, while the rate for bubble nucleation
approaches the finite nonzero value associated with the standard Kaluza-Klein
vacuum.
We will resolve this apparent paradox in detail below. The essential point
is that the magnetic field seen in four dimensions is not  uniquely
determined by the five dimensional solution.
{}For axisymmetric configurations, one must
choose an `internal space' by specifying a
Killing field with  closed orbits;
different choices yield different values of $B$. Physical
considerations restrict this choice so that $B$ is small compared
to the compactification
scale. For one
range of parameters and one choice of internal space, the Kerr instanton
yields a four dimensional solution with small $B$ which resembles the one
obtained by Witten. However, for another range of parameters, and a different
choice of internal space,  the Kerr instanton again yields a four dimensional
solution with small $B$ which now resembles a pair of accelerating monopoles.

A closer examination of these
decay processes contain further
surprises. First, as pointed out in \rafKK\ the ``bubble of nothing'' in five
dimensions
appears as a point-like singularity in four dimensions -- it does not
expand outward, instead space collapses in towards it.
In five dimensions, it turns out that the bubble wall
follows a geodesic, not a curve of uniform acceleration.
Second, in the pair creation of monopoles, the spacetime
between the monopoles dynamically decompactifies: the size of the fifth
direction increases with time, so the four dimensional description eventually
breaks down.

It has been suggested \witten\ that fundamental fermions could stabilize the
standard \KK\ vacuum. It appears that the same mechanism
eliminates bubble nucleation but
allows the pair creation of monopoles.

The outline of this paper is as follows. In the next section we discuss
the Kaluza-Klein solutions describing magnetic fields, and explain how a
given five dimensional solution can give rise to different four dimensional
descriptions. In section 3, we introduce the five dimensional
Kerr instanton, examine
its properties, and compute its action. Section 4 contains a review of
the \KK\ Ernst instanton, and establishes its equivalence to the Kerr instanton
of the previous section. The final section consists of a summary of our
results and the arguments as to how spinors can rule out bubble
formation but not pair creation.

\newsec{Uniform Magnetic {}Field}

In Einstein-Maxwell theory,  the closest analogue to a
uniform magnetic field is the Melvin spacetime
\melvin, which describes a static
cylindrically symmetric magnetic flux tube. The
generalisation of this solution to Kaluza-Klein
theory was constructed by Gibbons and Maeda \gm. It was
later realised that this solution can be obtained from
$M^5$ by simply identifying points in a nonstandard
way \refs{\dgkt, \dggh}. Explicitly,
the spacetime is given by the flat metric
in cylindrical coordinates
\eqn\kkmel{
ds^2=-dt^2+dz^2+d\rho^2+\rho^2  d\vp^2 +(dx^5)^2}
with the identifications
\eqn\idents{(t,z,\rho,\vp,x^5) \equiv
(t,z,\rho,\vp+2\pi n_1 R B +2\pi n_2,x^5+2\pi  n_1R)\ \ \ \forall n_1,n_2
\in Z\, .
}
The identification under shifts of
$2\pi n_2$ for $\vp$ and $2\pi  n_1R$ for $x^5$ are, of
course, standard. The new feature is that under a shift of $x^5$, one also
shifts $\vp$ by $2\pi n_1 R B$. Since $\vp$ is already periodic with period
$2\pi$, changing $B$ by a multiple of $1/R$ does not change the
identifications.
Inequivalent spacetimes are obtained only for $-1/2R < B \le 1/2R$.
More geometrically, one can obtain this spacetime by starting with \kkmel\
and identifying points along the closed orbits of the Killing vector
$l=\partial_5+B\partial_\vp$.

To obtain the four dimensional description, one must reduce along a Killing
field with closed orbits. An obvious candidate is $l$.
Introducing
the new coordinate $\tilde\vp=\vp-Bx^5$ which is constant along
the orbits of $l$, the metric becomes
\eqn\kkmelt{
ds^2=-dt^2+dz^2+d\rho^2+\rho^2 (d\tilde\vp+B dx^5)^2 +(dx^5)^2}
now with the points $(t,z,\rho,\tilde\vp,x^5)$ and
$(t,z,\rho,\tilde\vp+2\pi n_2,x^5+2\pi n_1 R )$
identified. In the new coordinates the Killing vector
is simply
$l=\partial_5$ and consequently it is straightforward
to perform the dimensional reduction.
We recast the metric in the following
canonical form
\eqn\fivmet{ ds^2 = e^{-4\phi/\sqrt 3}( dx^5 + 2A_\mu dx^\mu)^2
     +e^{2\phi/\sqrt 3} g_{\mu\nu} dx^\mu dx^\nu \  }
where $x^\mu$ are the four dimensional coordinates.
Note that with this decomposition into four dimensional fields
which do not depend on the fifth direction, the
five dimensional Einstein-Hilbert action up to surface terms becomes
\eqn\einact{
I = {1\over 16 \pi G_5} \int d^5x \sqrt{-{}^5\!g}\ {}^5\!R
= {1\over 16 \pi G_4} \int d^4x \sqrt{-g}\(R -2(\nabla\phi)^2
-e^{-2\sqrt{3} \phi} F^2 \)
}
where $G_5 = 2\pi R G_4$. We deduce that the unit of electric charge,
in these units, is $e = 2/R$.

In terms of four dimensional fields, \kkmelt\ is
\eqn\dmelv{
\eqalign{
&ds_4^2=\Lambda^{1/2}\left[-dt^2+d\rho^2+dz^2\right]
+\Lambda^{-1/2}\rho^2d\tilde\vp^2\cr
&e^{-{4\over{\sqrt{3}}}\phi}=\Lambda,\qquad
A_{\tilde\vp}={B \rho^2 \over 2\Lambda}\cr
&\qquad \Lambda=1+ B^2\rho^2\cr}
}
This solution describes a magnetic flux tube in the $z$ direction
and thus generalises the Melvin solution of Einstein-Maxwell theory.
The parameter $B$ gives the strength of the magnetic field on the axis
via $B^2 = \left.{1\over 2} {}F_{\mu\nu} F^{\mu\nu}\right|_{\rho = 0}$.

Although the choice of reducing to four dimensions along $l$ seems natural,
it is not unique. One could consider using the
Killing vector $\hat l=l +(n/R)\p_\vp$
for any integer $n$, which also has closed orbits.
The corresponding four dimensional solution is simply \dmelv, with
magnetic field parameter $B+n/R$. Recalling that the parameter $B$ in the
five dimensional metric is restricted to lie in the range $ -1/2R < B \le
1/2R$,
it would appear that
all values of the four dimensional magnetic field can be obtained.

However, we must consider the range of applicability of these spacetimes.
{}From \fivmet\ and \dmelv\ we see that for every $B \ne 0$,
the proper length of the circles in
the fifth direction grows linearly with $\rho$ for large $\rho$.
This seems to  cast doubt on their interpretation as  Kaluza-Klein backgrounds.
{}Fortunately, this is not a problem since physical
magnetic fields are not infinite in spatial extent.
We can
view \dmelv\ as an approximation to a constant
physical magnetic field
which is valid only for
$\rho <<1/|B|$, in which range three dimensional space
is approximately flat and the
internal circles have approximately constant length. (This is
not new to \KK\ theory: even in Einstein-Maxwell theory,
calculations of the decay of electromagnetic fields due to
pair creation of black holes use the same assumption
since the exact Melvin spacetime ``curls up'' far from the axis.)
In addition, in order for the fifth direction to remain unobservable, we
must consider length scales large compared to their size: $\rho >> R$.
Comparing these two restrictions on $\rho$ we see that there is a nontrivial
range of applicability only for $|B| << 1/R$. (If $R$ is of order the Planck
scale, this includes large magnetic fields in conventional units.)
In other words, if $|B| \sim 1/R$, the four dimensional metric
is curved on scales of order the compactification scale, so a four
dimensional interpretation is no longer appropriate.
Since the different four dimensional reductions change $B$ by multiples of
$1/R$, we see that for fixed $B$ at most one can be physically
reasonable.
Note that in contrast, due to the translational invariance,
there is no limit on the length of a physical flux tube
that can be well approximated by the Melvin solution.

\newsec{The Five Dimensional Kerr Instanton}
\subsec{The Geometry}

Myers and Perry have generalised
the four dimensional
Kerr solution
to
arbitrary dimensions $d\ge 4$ \myp.
{}For $d=5$, in addition to the mass, the solutions are labeled
by two angular momentum parameters. Asymptotically we can think
of these as describing a rotation in two orthogonal planes in $R^4$.
{}For our purposes we are interested in the case when only one
of the angular momentum parameters is non-zero. In this case
the Lorentzian metric is given by
\eqn\lkerr{
\eqalign{
ds^2=&-dt^2+\sin^2\theta(r^2+a^2)d\vp^2+{\mu\over \rho^2}(dt+a\sin^2\theta
d\vp)^2\cr
&+{\rho^2\over r^2+a^2-\mu}dr^2+\rho^2d\theta^2
+r^2\cos^2\theta d\psi^2\cr}}
where
$\rho^2=r^2+a^2\cos^2\theta$,
$\mu$ and $a$ are the mass and angular momentum
parameters and the range of the
angular variables is $0\le \theta\le \pi/2$, $0\le\vp\le 2\pi$,
$0\le \psi\le 2\pi$.

The instanton metric is obtained by setting $t=ix^5$ and $a=i\alpha$ with
$\alpha$ real:
\eqn\kinst{
\eqalign{
ds^2=&(dx^5)^2 + \sin^2\theta(r^2-\alpha^2)d\vp^2 - {\mu\over \rho^2}
(dx^5+\alpha \sin^2\theta
d\vp)^2\cr &+ {\rho^2\over r^2-\alpha ^2-\mu}dr^2+
\rho^2d\theta^2 + r^2\cos^2\theta d\psi^2\cr}}
where now
\eqn\row{
\rho^2=r^2-\alpha^2\cos^2\theta,}
This metric has a coordinate singularity at $r^2=r_H^2
\equiv\mu +\alpha^2$ (the
location of the black hole horizon in the Lorentzian metric). The potential
conical singularity can be
eliminated by a suitable periodic identification of the coordinates
$\vp$ and $x^5$.
To see this in detail, let us first introduce two quantities  encountered
in lorentzian black hole theory:
\eqn\avsg{
\Omega={\alpha\over\mu},\qquad \kappa={\sqrt{\mu+\alpha^2}\over\mu}}
where $\omega = i\Omega$ and $\kappa$ are the lorentzian angular
velocity and surface gravity, respectively, analytically continued to
imaginary values of the parameter $a$. The norm of the Killing vector
\eqn\kv{
l={\p\over \p x^5}+\Omega {\p\over \p \vp}}
consequently vanishes at $r=r_H$. Introducing the new
coordinate $\tilde\vp=\vp -\Omega x^5$, which is constant
along the orbits of $l$, we note that near
$r=r_H$, the metric \kinst\ can be written
\eqn\app{
\eqalign{ds^2\approx &(r-r_H)
f(\theta) (dx^5)^2 +(r-r_H) g(\theta) d\tilde\vp dx^5
+{\mu^2 \sin^2\theta\over(\alpha^2 \sin^2\theta+\mu)}
 d\tilde\vp^2\cr &+ {1\over 4\kappa^2}{f(\theta)\over (r-r_H)} dr^2 +\dots
\cr}}
where
\eqn\term{
f(\theta)={2 r_H (\alpha^2\sin^2\theta +\mu)\over \mu^2}
\qquad g(\theta)={4 r_H \alpha \sin^2\theta
(\alpha^2 \sin^2\theta+ 2\mu)\over \mu (\alpha^2 \sin^2\theta +\mu)}
}
and the elipsis denote terms that are not important for the following argument.
If at fixed $\tilde \vp$ we assume that $0\le x^5\le 2\pi/\kappa $ we
can introduce the coordinates
\eqn\xandy{
x=(r-r_H)^{1/2} \cos(x^5\kappa)\qquad y=(r-r_H)^{1/2} \sin(x^5\kappa)
}
The metric then takes the form
\eqn\mxy{
ds^2\approx  {f(\theta) \over\kappa^2}(dx^2 +dy^2)+ {g(\theta)\over \kappa}
(x dy- ydx)d\tilde\vp
+{\mu^2 \sin^2\theta\over (\alpha^2 \sin^2\theta +\mu)} d\tilde\vp^2+\dots}
which is clearly real and analytic at $r=r_H$ ($x=y=0$).
Thus the conical singularity is eliminated by requiring that
$x^5$ be periodic with period
$2\pi R$ at fixed $\tilde \vp$ where
\eqn\R{
R={1\over \kappa}={\mu\over \sqrt{\alpha^2+\mu}}
}
In terms of
of the $(x^5,\vp,r,\theta,\psi)$ coordinates we deduce that
the points $(x^5,\vp,r,\theta,\psi)$ and $(x^5+2\pi n_1 R,\vp+2\pi n_1
\Omega R+2\pi n_2,r,\theta,\psi)$ must be
identified\foot{Demanding that the metric is smooth on the axis
$g_{\vp\vp}=0$ we deduce that $\vp$ has period $2\pi$ at fixed $x^5$.}.

In the limit $r\to\infty$,  the instanton metric \kinst\ approaches
\eqn\mel{ds^2=(dx^5)^2+r^2
\sin^2\theta d\vp^2+dr^2 +r^2d \theta^2 +r^2\cos^2\theta d\psi^2
}
Using the cylindrical coordinates
$
\rho=r \sin\theta,z=r \cos\theta
$
with $z\ge 0$, we get
\eqn\mt{ds^2= (dx^5)^2 +\rho^2  d\vp^2 +d\rho^2+ dz^2+  z^2 d\psi^2
}
This is clearly flat.
Because of the identifications made on the angles $x^5,\vp$, we conclude
that asymptotically the instanton approaches
a euclidean Kaluza-Klein magnetic field with magnetic field strength
\eqn\bfield{
B=\Omega={\alpha\over \mu}}

Since $BR = \alpha/\sqrt{\alpha^2+ \mu}$, and $\a$ can take both positive
and negative values, we see that $B$ lies in the range $-1/R < B < 1/R$.
In section 2, we saw that for a uniform magnetic field, inequivalent
five-dimensional spacetimes were obtained only for $-1/2R < B \le  1/2R$.
This means that  the Kerr instanton with $B<-1/2R$ ($B>1/2R$) asymptotically
approaches exactly the same magnetic field solution as the instanton with
parameter $B + 1/R$ ($B-1/R$). This will be an important point
in the interpretation of the instanton as describing two modes of decay,
to be discussed shortly.

\subsec{The Euclidean Action}

In this section we calculate the euclidean
action of the above instantons. This will
be used in the next section when we interpret them as mediating a decay of
the Kaluza-Klein magnetic field.
The full euclidean action with boundary terms included
is only defined with respect to a reference background
and is given by
\eqn\eact{I = -{1\over 16\pi G_5} \int \sqrt g R - {1\over 8\pi G_5} \oint
 \sqrt h (K-K_0) }
where $K$ is the trace of the extrinsic curvature of the boundary, and
$K_0$ is the analogous quantity for the boundary embedded in the background
geometry. For the Kerr instanton, the appropriate background is the
(analytic continuation of the) Kaluza-Klein
magnetic field solution \kkmelt. Since our instanton is Ricci flat, the
only contribution to $S$ comes from the surface term. The metric induced
on the surface $r=$ constant in \kinst\ is
\eqn\surmet{
ds^2=\(1-{\mu\over r^2}\)(dx^5)^2 + \sin^2\theta\ (r^2-\alpha^2)d\vp^2
+\rho^2 d\theta^2 + r^2\cos^2\theta d\psi^2}
where we have only kept terms of order $O(1/r^2)$. Computing the derivative of
the volume element of this metric with respect to a unit radial vector yields
\eqn\extcur{ K\sqrt h = \sin(2\theta)\[ {3\over 2} r^2 - \mu -{\a^2 \over 4}
[3-\cos(2\theta) ] \]  }

The background contribution is easily computed using the fact that the
metric \kinst\ approaches flat space in the limit $\mu =0$ for all values of
$\a$ \myp. (This is reasonable since the total angular momentum is proportional
to $\mu$.)
It is clearly more convenient to use this representation of flat
space to embed the boundary isometrically, than the standard one.\foot{The
identifications needed to convert this flat space into the magnetic
field solution are identical to the ones with $\mu \ne 0$, and do not
affect local calculations such as the extrinsic curvature.}
Since $\mu$ only enters the metric \surmet\ in $g_{55}$, we can
embed it in flat space by taking a surface of constant $r$ in \kinst\ with
$\mu=0$, and letting the flat space coordinate $x^5$
have period $2\pi R (1-\mu/2r^2)$. We can then compute $K_0 \sqrt h$
and take the difference with \extcur\ to obtain
\eqn\diff{ (K-K_0)\sqrt h = -{\mu \sin(2\theta)\over 4}}
The euclidean action is therefore
\eqn\act{I_{\rm Kerr} = {\pi^2 \mu R \over 4 G_5}}
Using
\R\ and  \bfield, one finds
\eqn\actt{
I_{\rm Kerr}={\pi R^2\over 8 G_4} {1\over (1- R^2 B^2)}}

We can check the consistency of this result with the various
thermodynamic formulae for five-dimensional black holes
\refs{\myp}. The mass, $M$, and angular momentum, $J$, are
given by
\eqn\massang{M = {3\pi \over 8} {\mu \over G_5}\ ,\qquad
J = {2\over 3} M a
}
where $a$ is the Lorentzian rotation parameter. The Smarr relation
is
\eqn\smarr{ M = {3\over 2} \(TS + \omega J\)
}
with the entropy given by $S = {1\over 4 G_5} A_H$, $A_H$ being the
3-area of the horizon. The thermodynamic
potential $W$ is
\eqn\thermpot{W = M - TS - \omega J = {1\over 3} M
}
and thus the euclidean action is
\eqn\easyact{ I_{\rm Kerr} = {W\over T} = {1\over 3} {M\over T} = 2\pi R
{M\over 3}
}
which agrees with our direct calculation.
Note that we can obtain the action of five-dimensional
Schwarzschild as the limit of $I_{\rm Kerr}$ for
zero $B$: $I_{\rm Schw} = {\pi R^2 \over 8 G_4}$. This differs by
a factor of two from the result of Witten \refs{\witten}.

\subsec{Interpretation as a Decay}

In order to show that the instanton \kinst\ describes the semi-classical
instability of a Kaluza-Klein magnetic field, it suffices to
find a surface of zero extrinsic curvature (zero momentum). One can then use
the
fields on this surface as initial data to obtain a real lorentzian metric
which describes the spacetime into which the static magnetic field decays.

Such a zero-momentum surface in \kinst\ is easy to find, and is given
by  $\psi =$ constant. In fact, to obtain a surface which is complete,
we need to take both
$\psi = 0$ and $\pi$, since $\psi$ is an angular coordinate with
regular origin at $\theta = \pi/2$. The induced metric on this surface
turns out to be just the
four dimensional
euclidean Kerr-Newman metric with zero mass.  The two
surfaces $\psi = {0,\pi}$ each have $0\le \theta \le \pi/2$ and cover half
of the space. But they join at $\theta = \pi/2$, and the full
zero-momentum slice is
conveniently represented by letting $\theta$ take its usual range
$0\le \theta \le \pi$.

The lorentzian evolution of this initial data
is obtained from \kinst\ by
rotating the coordinate $\psi$, $\psi\to it$:
\eqn\linst{
\eqalign{
ds^2=&(dx^5)^2+\sin^2\theta(r^2-\alpha^2)d\vp^2-{\mu\over \rho^2}
(dx^5+\alpha \sin^2\theta
d\vp)^2\cr &+{\rho^2\over r^2-\alpha ^2-\mu}dr^2+\rho^2d\theta^2
- r^2\cos^2\theta dt^2\cr}}
To understand what this metric represents, let us first set
$\a =0$. The metric \linst\ then reduces to
\eqn\schwi{ ds^2 = \(1-{\mu \over r^2}\) (dx^5)^2
 + \(1-{\mu \over r^2}\)^{-1} dr^2
  + r^2[-\cos^2 \theta dt^2 +  d\theta^2 + \sin^2 \theta d\vp^2 ]}
This is precisely the solution found by Witten in his study of the decay of
the Kaluza-Klein vacuum.
Witten presented the solution in a different set of
coordinates
\eqn\wischw{ ds^2 = \(1-{\mu \over r^2}\) (dx^5)^2
 + \(1-{\mu \over r^2}\)^{-1} dr^2
   +r^2[-d\tilde t^2 + \cosh^2 \tilde t (d\tilde \theta^2 + \sin^2 \tilde
\theta d\vp^2)] }
Both of these metrics can be obtained by starting with the five-dimensional
euclidean Schwarzschild solution, and analytically continuing the round metric
on the $S^3$.
If one starts with
$ds^2 = d\chi^2 + \sin^2 \chi d\Omega_2$ and sets $\chi = (\pi/2) + i\tilde t$
one
obtains the form \wischw. The metric in brackets is just three dimensional
de Sitter space. If instead one starts with $ ds^2=
d\theta^2 + \sin^2 \theta d\vp^2 + \cos^2 \theta d\psi^2$ and sets $\psi = it$,
one obtains the form \schwi. The metric in brackets is again
three dimensional de Sitter space, but now in static coordinates.
These do not cover the entire spacetime, but only the region inside the horizon
at $\theta = \pi/2$. Note that the initial 4-surfaces $t =0$  in the two
metrics
are identical. Because $\p/\p x^5$ vanishes at $r=r_H$ the initial spacelike
surface is spherically symmetric and has topology $R^2\times S^2$.
The 2-surface $r=r_H$ is a totally geodesic (and hence minimal) submanifold
of the initial 4-surface with area $4\pi\mu$. The subsequent evolution is
easier
to see in the metric \wischw\ which covers the entire 5-dimensional spacetime.
The
surface $r=r_H$
expands outwards with area increasing like ${\rm cosh}^2t$. This is Witten's
expanding
bubble. The isometry group is $U(1) \times SO(3,1)$.

Returning to the general solution \linst, we see that at $t=0$, the surfaces
of constant $r$ and $x^5$ again have a minimum area of $4\pi \mu$ which
is obtained when $r=r_H$. Even though we no longer have spherical symmetry,
this surface is geometrically singled out since it is
a totally geodesic two-sphere.
Although the metric appears static, it is directly
analogous to \schwi. The time translation symmetry is a boost, since
for large $r$ the metric approaches the analytic
continuation of \mt
\eqn\amt{ds^2= (dx^5)^2 +\rho^2  d\vp^2 +d\rho^2+ dz^2- z^2 dt^2.}
One can again introduce the coordinates ($\tilde t,\tilde \theta$)
used in \wischw\ which allow one to extend through the coordinate singularity
at $\theta=\pi/2$ (which corresponds to a Killing horizon of
the boost Killing vector field $\p/\p t$). Although these coordinates
cover the entire spacetime, the  $\p/\p\tilde t$ vector field is not
hypersurface
orthogonal when $\alpha\ne 0$. Nevertheless, we can still conclude that
the ``bubble" $r=r_H$ is a deformed version of the expanding three dimensional
de Sitter metric.


As we mentioned earlier, there are two distinct Kerr
instantons that asymptotically approach a given five-dimensional magnetic field
solution, \kkmel\ with $|B_0| \le 1/2R$. The obvious one has $\a/\mu = B_0$
while the less obvious one, ``shifted Kerr", has
$\a/\mu = B_0 \pm 1/R$ (where the upper sign is chosen when $B_0$ is negative
and the lower sign when $B_0$ is positive).
Thus there are two separate decay modes; the
one that dominates will be the one with the lowest action.
{}From \actt\ we
see that if $|B| \equiv |\a/\mu| < 1/2R$ the first will dominate while if
$|B| \equiv |\a/\mu|>1/2R$ the second
will. However, we argued in section 2, that these solutions are physically
reasonable only if $|B| << 1/R$.  Thus, the first instanton is physically
the most important. Since even this instanton has a larger action
than the one with
$B=0$, we see that the presence of a magnetic field tends to suppress
the decay of the vacuum.
We have plotted the action of the two instantons in figure 1.

It should be emphasised that the two instantons are the same five
dimensional Kerr instanton but with different values of its parameters. On the
other hand we will see in the
next section that the more physical four dimensional
interpretations differ substantially.
\ifig{\Fig\actions}
{Actions of two instantons mediating the
decay of a Kaluza-Klein magnetic field  versus
magnetic field
strength. The solid line is the Kerr instanton and the dotted
line is the ``shifted Kerr" instanton. The range of inequivalent magnetic
fields is given by $-1/2R<B\le 1/2R$ and the ``physical'' range
is
close to the $I$ axis. Hence the ``unshifted'' decay dominates.
}{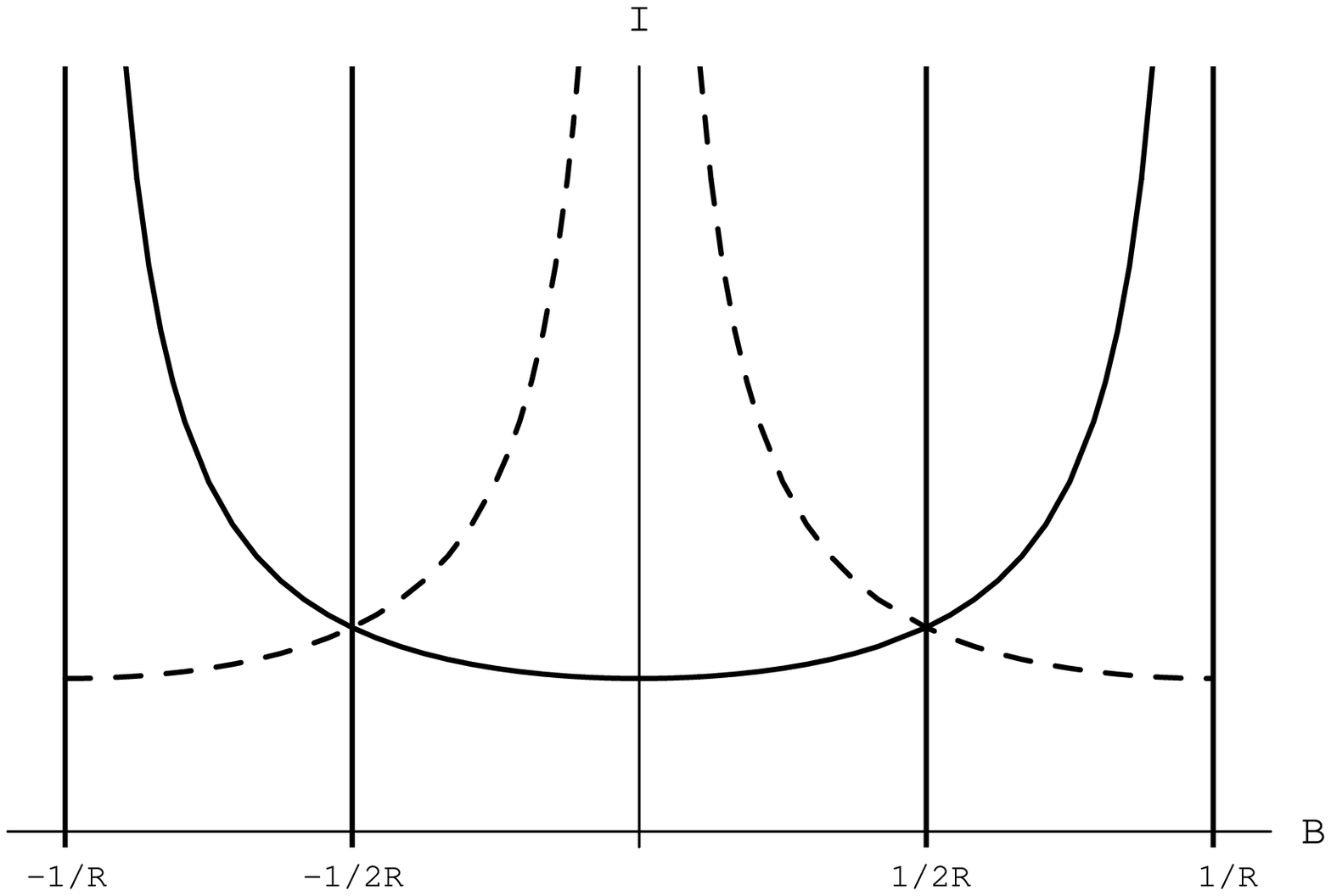}{6.0}

\subsec{Four-Dimensional Description}

We now wish to examine what the five-dimensional
lorentzian solution
\linst\ looks like in terms of four-dimensional fields: i.e. to
relate it to physics. As we discussed
in section 2, this requires a choice of Killing field $l$ with closed orbits
and the issues raised there regarding the physical justification
of the Kaluza-Klein reduction will
be relevant.
If we use coordinates in which the Killing field is simply $l = \p / \p x^5$,
then the four-dimensional
fields can be read off after
writing the five-dimensional
metric in the form \fivmet.

Let us start with the case $\a =0$
\schwi. If we reduce along the symmetry $l=\p / \p x^5$, there is no
four-dimensional
Maxwell field, and the four-dimensional
metric is
\eqn\wifd{ ds^2_4 = \(1-{\mu \over r^2}\)^{1/2}\[\(1-{\mu \over r^2}\)^{-1}
dr^2
   +r^2(-\cos^2 \theta dt^2 + d\theta^2 + \sin^2 \theta d\vp^2) \]}
In this metric, the ``bubble'' at $r=r_H$ has zero area and is a point-like
singularity
\foot{Although we refer, here and subsequently, to this as a
singularity in four dimensions, it should be noted that we
are using this as short hand for the statement that
it is a point in whose neighbourhood
the four dimensional
description breaks down and the true five dimensional
nature of the spacetime, which is completely regular, necessarily
manifests itself. It is not a singularity in the
sense of a breakdown of the physics. The same comment applies to
discussions of \KK\ monopoles.}
 which is timelike \rafKK. However this four dimensional spacetime
differs from other
static spacetimes with naked singularities such as the negative mass
Schwarzschild solution. The reason is that the timelike symmetry is a boost,
so the singularity  is following the orbit of a boost and hits null infinity.
More physically, one could view the singularity as being ``at rest",
with space ``falling into it".
Both future and past null infinity are incomplete.

If we reduce along the symmetry $\hat l = \p_5 + (n/R) \p_\vp$ then the
situation is different. To see this, we introduce the new coordinate
$\tilde \vp = \vp - (n/R) x^5$ which is constant along $\hat l$. Since we
have singled out one rotation direction, the four-dimensional
spacetime will no
longer have the full $SO(3,1)$ symmetry, but instead will have only a time
translation and $U(1)$ symmetry. In the new coordinates
\eqn\gff{g_{55} = \(1-{\mu \over r^2}\) +{n^2 r^2 \over R^2} \sin^2\theta}
$$ g_{5 \tilde \vp} = {n r^2 \over R}\sin^2\theta $$
Notice that $g_{55}$ no longer vanishes everywhere on the horizon but
only at the poles $\theta = {0,\pi}$. More geometrically, the Killing vector
field
$\hat l$ has a ``nut" and an ``antinut" at the north and south pole,
respectively.
In the case $n^2=1$ these
nuts are self dual (anti-self dual) in the sense of \gh\
and correspond to monopoles (antimonopoles), as we will see in detail
in section 4 in the context of the Kerr instanton. The four-dimensional
gauge field, $A_{\tilde \vp} = g_{5 \tilde \vp}/(2 g_{55})$,
is now nonzero
and asymptotically
describes a uniform magnetic field with strength $B=n/R$.
This means that these four-dimensional reductions of Schwarzschild are
in some sense unphysical: as discussed in Section 2, the \KK\
reduction only makes sense when the
four-dimensional magnetic field strength, $B$, satisfies
$B<<1/R$. In other words we, from our four-dimensional
point of
view, would never see magnetic fields of strength $n/R$ and
the question of how we would see them decay is moot. We will however
continue to analyse these reductions as a simpler exercise before
looking at the four-dimensional reductions of Kerr, some of which
{\it will} be physically relevant.

Using \fivmet\ and \gff, the four-dimensional
metric is given by
\eqn\fdsch{ ds^2 = g_{55}^{1/2} \[- r^2\cos^2\theta dt^2 +
  \(1-{\mu \over r^2}\)^{-1} dr^2
+r^2 d\theta^2 + {(r^2 - \mu)\sin^2 \theta \over  g_{55} }d\tp^2\]}
The important point is that $g_{\tp\tp} = 0 $ at the horizon while $g_{\theta
\theta}$ remains nonzero. Thus the horizon is no longer a point, but rather
a line. The endpoints of this line $\theta = {0,\pi}$
are curvature singularities, but away from these points the metric near
this line (on a $t =$ constant surface) is simply proportional to
\eqn\linemet{ dz^2 + d\r^2 + {\r^2 \over n^2 }  d\tp^2}
where we have set $z= r_H \theta$, $\r^2 = r^2 - \mu$, and used the fact
\R\ that  $R^2 = \mu$.
Thus for $n = \pm 1$ the line is completely smooth, while for  $|n| >1$ there
is a conical singularity. Since the deficit angle is positive, the conical
singularity represents a string connecting the two singularities.
Given the boost symmetry of the five-dimensional
metric, it is clear that under time evolution, the two pointlike singularities
will expand away from each other and hit null infinity, for all values of
$n$.

We now consider the general case \linst\ with $\a \ne 0$. The natural choice
of Killing field to reduce along is $l = \p_5 +\Omega \p_\vp$
which vanishes at the horizon. Since $l=0$ at the horizon, it follows that
the four-dimensional
metric will be singular there. Since the boost symmetry
is preserved under dimensional reduction, we see that the four-dimensional
metric resembles the first reduction of $\a=0$ discussed above. There is
a single naked singularity and null infinity is incomplete. The
asymptotic value of the four-dimensional magnetic field is
$\Omega=\alpha/\mu$ and so we can only interpret this as the
four-dimensional view of the decay if $|\alpha/\mu| << 1/R$.

More generally, we can reduce along the Killing field $\hat l =
l +(n/R) \p_\vp$ for any integer $n$. As before, this is accomplished by
introducing the new coordinate $\tilde \vp = \vp - B x^5$ where $B = (\a/\mu)
+(n/R)$. We then find
\eqn\gffa{ g_{55} = 1-{\mu\over \rho^2} (1+\a B\sin^2\theta)^2
     + B^2 (r^2 - \a^2) \sin^2\theta  }
$$ g_{5\tilde \vp} = B(r^2 - \a^2) \sin^2\theta - {\mu \a \sin^2\theta
 \over \rho^2} (1+\a B\sin^2\theta)  $$
 One can show that $g_{55} = 0 $ on the horizon for $B = \a/\mu$ as expected,
 but is nonzero on the horizon for other values of $B$ (except at $\theta =
 0,\pi$).
 {}From this, and \fivmet, we can compute the four-dimensional
metric.
\eqn\fdkerr{ ds^2 = g_{55}^{1/2} \[- r^2\cos^2\theta dt^2 +
{\rho^2 \over r^2 -\a^2 -\mu} dr^2
+\rho^2 d\theta^2 + {(r^2 -\a^2 - \mu)\sin^2 \theta \over g_{55}} d\tp^2\]}
We see that $g_{\tp\tp}$ again vanishes on the horizon so that the horizon
is now a line which ends in two naked singularities. Those
singularities, due to the boost symmetry, accelerate away to infinity.
{}Furthermore, on the horizon
\eqn\gffhor{ \rho^2 g_{55} = n^2 r_H^2 \sin^2\theta}
so if we set $\r^2 = r^2 - \a^2 -\mu$, the metric is again proportional
to \linemet\
which is regular for $|n| = 1$.

As before, we can only interpret these four-dimensional descriptions
sensibly as \KK\ reductions when the four dimensional magnetic field is
much less than $1/R$. However, unlike the Schwarzschild case, this
condition can now be satisfied even with $n \ne 0$: We need
$|(\alpha/\mu) + (n/R)|<< 1/R$ which (since $|\alpha/\mu| <
1/R$ from \R)
will hold if either $\alpha/\mu$ is positive, close to $1/R$ and $n=-1$,
or
$\alpha/\mu$ is negative, close to $-1/R$ and $n=+1$.

To summarize, we have seen that there are two Kerr instantons which
asymptotically approach a given five-dimensional
Kaluza-Klein magnetic field.
If we
start with a four-dimensional magnetic field with strength
$0 \le B << 1/R$, then we can either use the Kerr instanton \kinst\
with $\a/\mu = B$ and reduce along $l = \p_5 + B\p_\vp$,
or take the Kerr instanton with $\a/\mu = B -1/R$ (``shifted Kerr")
and reduce along
the Killing field $\hat l = l + (1/R) \p_\vp$.
(Similarly for small
negative $B$, there are two decay modes.) In the first case,
a single naked singularity appears in space, while in the second, there is
a pair of naked singularities accelerating away from each other.
For
small $B$ the second process is highly suppressed with respect to
the first. The actions for these two instantons are given in figure 1
by the solid and dotted lines, respectively.


\newsec{Kerr is Ernst in Kaluza-Klein}

The four-dimensional picture of two objects
accelerating away from each other in a magnetic field is
reminiscent of another known
solution in \KK\ theory: the \KK\
Ernst solution \refs{\dgkt, \dggh}.
In this section we will prove that the
five-dimensional Kerr instanton and the extremal Kaluza-Klein
Ernst instanton are actually the same.

\subsec{Review of \KK\ Ernst: Pair Creation of Monopoles}

In Einstein-Maxwell theory, a Melvin magnetic solution can decay via
the pair production of
(extremal and non-extremal)
charged black holes \refs{\gwg,\garstrom,\dggh}. The
instanton for this process is the euclidean section of a solution found
by Ernst \refs{\ernst} which describes charged black holes
accelerating in a magnetic field.
Similarly, a Kaluza Klein magnetic field can also decay
via the pair creation of
Kaluza Klein monopole-anti-monopole
pairs \refs{\dggh}. The instanton for this process is
the euclideanisation of a solution that describes a Kaluza-Klein
monopole and anti-monopole
accelerating away from each other in a Melvin
background\foot{The form of the solution given here differs
slightly from that in the references. Here we have chosen coordinates
such that $\Lambda \rightarrow 1$ at infinity.
This turns out to be much more convenient,
and this new form of the
solution can also be taken for all values of the
dilaton coupling $a$ (here, $a= \sqrt{3}$). It is
obtained by the same generating transformation used in
\refs{\dgkt} but starting with a form of the C-metric in which the
gauge potential vanishes on $x=\xi_3$.
The formulae in
\dggh\ may be used here if some care is taken in making
the substitutions $\Lambda(\xi_3) \rightarrow 1$
and $\Lambda(\xi_4) \rightarrow
(1 + {1\over 2}(1 + a^2) bq(\xi_4-\xi_3))^2$.}
\refs{\dgkt, \dggh}
\eqn\final{\eqalign{
ds^2 =& {
\Lambda  f(y) \over  f(x) }\( dx^5 + 2A_\Phi d\Phi\)^2\cr +
  {1
\over A^2 (x-y)^2} &\[ -f(x)^2\({g (y) d\tau^2 \over f (y)}
 +{dy^2 \over g(y)} \)  + f (y)\( {f (x) dx^2 \over g (x)}
  +{g (x) d\Phi^2 \over \Lambda} \) \right] \cr
}}
where
\eqn\kkla{\eqalign{
A_\Phi=&-{1\over 2b\Lambda}(1+2bq(x-\xi_3))
+{1\over 2b}\cr
\Lambda=&(1+ 2bq(x-\xi_3))^2+
{b^2 g(x)f(x)\over A^2(x-y)^2}\cr
 f(\xi)=&(1+r_-A\xi)\cr
 g(\xi)=&\left[1-\xi^2-r_+A\xi^3\right]\cr
 4 q^2 =& r_+ r_- \, .\cr
 }}
The roots of $g(\xi)$ are $\xi_2<\xi_3<\xi_4$.
The surface $y = \xi_3$ is the acceleration horizon and the
coordinates are restricted by $\xi_2\le y \le \xi_3$ and $\xi_3
\le x \le \xi_4$.
$\tau$ is a euclidean time coordinate whose period,
$\Delta\tau$, is chosen so
that the acceleration horizon is regular.
The
zero-momentum slice is $\tau = 0, {1\over 2} \Delta \tau$.
$x^5$ is also periodic with period $2\pi R$.
As infinity is approached, $x, y \rightarrow \xi_3$, the solution
tends to the Melvin spacetime \dmelv, euclideanised,
with magnetic field parameter
$\B \equiv b\nu$, where we have defined $\nu \equiv  {1\over 2} g'(\xi_3)
f(\xi_3)^{-{1\over 2}}>0 $ for convenience.

The five parameters, $R$, $b$, $r_+>0$, $r_->0$ and $A>0$ are restricted
by three conditions. The first requirement is that the
root of $f(\xi)$, $\xi_1$,
 be equal to the lowest root of $g(\xi)$, $\xi_2$. If this
condition does not hold then  the solution describes non-extremal,
magnetically charged black holes, not monopoles.
The second condition is
\eqn\nonodal{-
g'(\xi_3)f(\xi_3)^{-{1\over 2}}
 =
g'(\xi_4)f(\xi_4)^{-{1\over 2}}
\Lambda(\xi_4)^{-{1\over 2}}\, .
}
which is needed to enforce regularity of the solution on the axis of
symmetry. The range of the azimuthal angle, $\Phi$, is
$\Delta\Phi = {2\pi\over \nu}$.
The condition \nonodal\ ensures that
choosing $\Delta\Phi$ as the range of $\Phi$ eliminates
conical singularities at both $x=\xi_3$ and $\xi_4$.
In four dimensions, in the weak field  limit,
this condition is physically transparent, it is simply
Newton's law: $mA = qb$, where for weak fields we can identify
$m,A,q$ with the mass, acceleration and charge of the black holes and $\hat
B\approx b$.

The final condition on the parameters is
given by the geometrical
analogue of the Dirac quantisation condition in the
presence of magnetic charge,
which is
indeed its four-dimensional manifestation.
We can  reduce \final\ to
four dimensions along $\p_5$,
calculate the physical
magnetic charge  of the monopole, $\q$ which
must be an integer multiple of $R/4$ to
eliminate conical singularities at the poles in five dimensions:
\eqn\diracq{ \q \equiv q
{
 (\xi_4-\xi_3) \over 2 \nu
\(1+ 2 qb (\xi_4-\xi_3)\)} = k {R\over 4}
}
where $k$ is an integer. Since the unit of electric charge
is $e=2/R$ we see that this gives us the Dirac quantisation
condition.
At the center of the monopole, $y=\xi_2$, the solution \final\
approaches that of the static \KK\ monopole of charge $\q$
\refs{\dggh}. Thus,
when $k =\pm 1$ this corresponds to the
Hopf fibration of $S^3$ and
the spacetime is completely regular at the
origin, whereas
for $|k| > 1$, the higher Hopf fibrations,
there is an orbifold singularity. We
therefore restrict attention to the cases $k=\pm 1$.
After imposing these conditions the independent parameters
can be chosen to be $\hat B$ and  $R$.

\subsec{The Action}
 The action of \final\ is \refs{\dggh}

\eqn\monact{I_{\rm mon} =
{2\pi \q^2\over G_4} {\Lambda(\xi_4)}{(\xi_3-\xi_2)\over
(\xi_4-\xi_3)}
}
It is possible to express the action in terms of the physical
magnetic field and monopole charge.
We have
\eqn\xifour{
\Lambda(\xi_4) = \(1 + 2 b q (\xi_4-\xi_3)\)^2
}
and condition \nonodal\ gives us
\eqn\nns{\Lambda(\xi_4) = {\xi_4 - \xi_2 \over \xi_3 - \xi_2 }
}
Then,
\eqn\working{
{\Lambda(\xi_4) (\xi_3-\xi_2)\over
(\xi_4-\xi_3)} = { \xi_4-\xi_2 \over \xi_4-\xi_3} =
\( 1 - \Lambda(\xi_4)^{-1} \)^{-1}
}
 {}From \diracq\  and the definition of $\B$ we have
\eqn\nexxt{ \Lambda(\xi_4) = \(1 - 4 \B\q \)^{-2}
}
and finally we obtain\foot{We can do a similar calculation for the
action of the instanton for pair creation of extremal
black holes for all values of the dilaton coupling $a$ \refs{\dggh}, expressing
it in terms of the physical magnetic field and charge:
$$
I_{\rm ext}={2 \pi{\q}^2\over G_4}
{ 1 \over {1 - \( 1 - (1+a^2) \B \q \)^2}}
$$}
\eqn\actnew{I_{\rm mon}(\B) =
{2 \pi{\q}^2\over G_4} { 1 \over {1 - \( 1 - 4 \B \q \)^2}}
= {\pi R^2 \over 8 G_4} {1 \over {1 - \(1 - |\B|R\)^2}}
}
where in the last step we have used ${\hat q}{\hat B}\ge 0$ which follows from
\nonodal\ and \diracq.


Comparing with \actt,  we see that the actions are
equal up to a shift in the magnetic field parameter by an amount
$1/R$: thus the action for the Ernst instanton is given
precisely by the dotted
line in figure 1.

\subsec{The Equivalence}

The actions of \kinst\ and \final\ indicate that we should look
for a coordinate transformation between \kinst\ with $\alpha > 0$ and
$B = {\alpha\over\mu}$ and \final\ with $\B = b \nu = B - 1/R < 0$.
This value of $b$ requires, by \nonodal\ and \diracq\ that
$q<0$ and
\eqn\qneg{ \q \equiv q
{
 (\xi_4-\xi_3) \over 2\nu
\(1+ 2 qb (\xi_4-\xi_3)\)} = - {R\over 4}
}
and from \nns
\eqn\qbpos{
1 + 2 q b (\xi_4 - \xi_3) =  \sqrt{{(\xi_4 - \xi_2)}\over{(\xi_3
-\xi_2)}}
}
which, together with the definition $\nu$, and \R\ and
\bfield\ imply
\eqn\mualpha{
\eqalign{
\mu =& {R^2 \over 1 - B^2 R^2} = {4 q^2 \over \nu^2} (\xi_3 -\xi_2)
        (\xi_4-\xi_3)\cr
\alpha =& {BR^2 \over 1 - B^2 R^2} = -{2q \over \nu} (\xi_3 - \xi_2)\, .\cr
}}

Next we note that $g_{55}$ in both cases tends to 1 at infinity,
suggesting we take the two $x^5$ coordinates to be equal.
However,
the cross term between $x^5$ and $\vp$ in \kinst\ tends to zero at
infinity whereas the cross term between $x^5$ and $\Phi$ in \final\ gives rise
to a Melvin magnetic field at infinity of strength $B-1/R$. We conclude
that in order to compare the two solutions we must change coordinates
in \final\ (we could choose to change coordinate in \kinst\ but
that turns out to be more complicated) so that the cross-term between
the new azimuthal coordinate and $x^5$ vanishes at infinity. This is
achieved by setting
\eqn\coordt{\Phi = \vp' - {1\over \nu}(B-1/R)x^5 }
and we obtain
\eqn\kkc{
\eqalign{
ds^2 =& {
f(y) \over  f(x) }\( dx^5 + 2A'_{\vp'} d\vp'\)^2\cr +
  {1
\over A^2 (x-y)^2} &\[  -f(x)^2\({g (y) d\tau^2 \over f (y)}
  +{dy^2 \over g(y)} \)  + f (y)\( {f (x) dx^2 \over g (x)}
  +g (x) d{\vp'}^2 \) \right] \cr
}}
where $A'_{\vp'} = q(x - \xi_3)$. This is locally the \KK\ C-metric
\refs{\dgkt}. However, it differs in that the identifications on the
coordinates
$\phi$ and $x^5$ are similar to that of Kerr as given after \R\
and so it still asymptotically approaches a Kaluza-Klein magnetic field
with strength $B$.
The fact that the coordinate transformation
\coordt\ results in the C-metric is an immediate result of the
observation that the Ernst solution is obtained from the
C-metric by just the reverse of this transformation \refs{\dgkt}.

One can now verify that \kinst\ can be transformed into \kkc\ by
the coordinate identifications:
\eqn\change{
\eqalign{
 \vp =& \nu \vp'\cr
 \psi =& \nu \tau \cr
 r^2 =& -\mu {(y - \xi_4)(x-\xi_2) \over (\xi_4-\xi_3)(x - y)} \cr
 \cos^2{\theta} =& - {(y - \xi_3)(x-\xi_2) \over (\xi_3-\xi_2)(x - y)} \cr
}}

It follows that the lorentzian Ernst solution,
\final\ with $\tau= it'$,  and the doubly
continued lorentzian Kerr solution \linst\ are the
same.
Thus the five-dimensional
solution which was previously interpreted as describing
two Kaluza-Klein monopoles accelerating
apart in a magnetic field, is in fact the same as the one describing
an expanding bubble.
Contrary to one's expectation,
the five-dimensional
space does not have two localized regions of
curvature. These appear in four dimensions as a result of
the reduction. In fact,
the two singularities that appear in
the ``shifted'' reduction of the Kerr solution are now revealed
to be none other than a \KK\  monopole and anti-monopole with
charges  $\pm R/4$.

The surprising equivalence between these two solutions
raises a number of issues which we now address.
In \dggh\ it was shown that the centers of the monopoles in
the extreme Kaluza-Klein
Ernst solution are not really accelerating, but in fact
follow geodesics in five dimensions.
How is this compatible with the fact that this
spacetime is equivalent to an expanding bubble? Since the worldlines
for the monopole centers are the North and South poles of the bubble,
this is consistent only if the bubble itself is not accelerating! To
confirm this, consider the bubble \wischw\ obtained from the five
dimensional
Schwarzschild solution. The worldline corresponding to
constant $r, \theta, \vp$ has tangent vector $u = (1/r) (\p /\p t)$. The
acceleration of this worldline is $A_\nu = (1/r) \nabla_\nu r$ whose
norm
vanishes as one approaches the bubble at $r^2 = \mu$. Thus, a
five-dimensional observer
near the bubble does not have to undergo large acceleration to stay away
from the bubble. A more general argument that the bubble does not
accelerate (which applies to Kerr as well) is simply that it is the fixed point
set of a continuous isometry and therefore must be totally geodesic
\gh.
In four dimensions however, observers do need to
accelerate more and more to stay away from the singularity.

One can clearly take the angular momentum parameter to zero in the Kerr
solution and obtain the Schwarzschild metric.
What is the analog of this limit for
the Ernst solution? {}From equation \mualpha\ we have
\eqn\swern{ \mu = {\alpha^2 \over (\xi_3 - \xi_2)} (\xi_4-\xi_3)  }
Since $\xi_4 - \xi_3$ is always finite, we see that the limit $\a \to 0$
with $\mu$ fixed corresponds to the limit $\xi_2 \to \xi_3$. Since the
range of $y$ in the Ernst instanton is $\xi_2 \le y \le \xi_3$, this
is clearly a singular limit. To obtain a regular limiting geometry,
one has to also rescale the coordinates $x$ and $y$. The appropriate rescalings
can be derived from \change\ since they just correspond to keeping $r$ and
$\theta$ finite and nonzero. The result is a description of the Schwarzschild
metric in accelerating coordinates.
In some sense the usual Ernst coordinates
include a factor of the Kerr angular momentum which must be removed before
taking the Schwarzschild limit.

In a similar vein, one might ask what is the analog of the extreme
Kerr solution. One can see from \lkerr\ that in five dimensions, the
lorentzian Kerr
solution never has a degenerate horizon. If we make the angular momentum
parameter too large, the horizon becomes singular. However, the analytically
continued Kerr instanton \kinst\ is regular for all values of $\mu$ and
$\a$ and thus there is no analog of the extremal limit. Of course,
the extreme Ernst
solution is itself the limit of a more general non-extreme solution. It
follows that there is a deformation of the five-dimensional
Kerr metric  under which it
describes two non-extreme Kaluza-Klein black holes accelerating
apart. To obtain it one can, for example, substitute in the non-extremal
C-metric $x$ and $y$ as functions of $r$ and $\theta$ given by \change.

We have seen that the Kaluza-Klein Ernst solution can be rewritten in a simpler
way as the Kerr solution. It is thus natural to ask whether
the original Ernst solution in Einstein-Maxwell theory can similarly
be rewritten in a simpler form.  More generally, consider
the one parameter family of theories with metric, Maxwell field, and dilaton
where the parameter $a$ governs the coupling between the dilaton and Maxwell
field. There is an analog of the Ernst solution for each value of the
parameter $a$ \dgkt.
Kaluza-Klein theory corresponds to $a^2 = 3$.
One can, for example, utilise the coordinate transformation \change\ to obtain
another
form of these metrics for $a^2 \ne 3$, perhaps leading to new insights.

In \dggh, the topology of the Kaluza-Klein Ernst instanton was shown to
be $S^5$ with an $S^1$ removed. How does this compare with the topology
of the Kerr instanton? The Kerr instanton has the topology of a
five-dimensional euclidean black hole: $R^2 \times S^3$ (with metrically the
$R^2$ in the
shape of a cigar). But $S^5 - S^1$
is equivalent to $R^4$ with a line removed, which is indeed $R^2 \times S^3$.
So the spacetimes are equivalent globally, and not just locally.
The lorentzian analog of
this statement
explains how the positive energy theorem is violated.
The extremal Ernst solution has zero total mass since there is a boost symmetry
but it is clearly not the Melvin background. However,
as first pointed out by Witten \witten, the positive mass theorem holds in
Kaluza-Klein theory only if the manifold is globally of the form $M\times S^1$,
for some four manifold $M$.

The topology of the spatial slices, including the zero-momentum slice,
of the Kaluza-Klein Ernst-Kerr solution is $S^4 - S^1 \cong R^2 \times S^2$.
We can argue that this is the  spatial topology of {\it any} monopole
anti-monopole configuration, for example one which is asymptotically
the \KK\ vacuum rather than Melvin, as follows. The
topology of a monopole-anti-monopole configuration
can be described generally
as the union $A\cup B\cup C$
where
 $A$ and $B$ are both  four balls $D^4$ corresponding to the monopole and
antimonopole and
$C$ is the non-trivial $U(1)$ bundle over $R^3\#D^3\#D^3$ ($R^3$ with two three
balls removed)
which has zero winding over the sphere at infinity, and windings $+1$ and $-1$
over
the other two $S^2$ boundaries \rafff. This description fixes the topology
uniquely and is independent
of whether the metric tends to the vacuum or a magnetic field
at infinity. Thus the topology must be $ R^2 \times S^2$, since we
know that is the topology of
the pole-anti-pole configuration in  Kaluza-Klein Ernst-Kerr.

A final interesting observation is that
after two \KK\ monopoles are created
and accelerate away to infinity, the
spacetime dynamically decompactifies. This is most easily seen using the
Ernst form of the metric \final.
The coordinate $y$ becomes timelike for $y> \xi_3$, and the late
time behavior corresponds to fixing  $x$ ($\ne \xi_3$) and all coordinates
other than
$y$, and then
taking the limit $y \to x$. It is easy to see that in this limit,
$g_{55}$ diverges.
In other
words the fifth dimension decompactifies and the four-dimensional
reduction is no longer valid. We should note that this is the
case at the level of the whole solution and since we are only considering
the solution close to the axis,
as an approximation to the decay of a real magnetic
field, its physical significance is unclear.

\newsec{Discussion}

To summarise, we have seen that magnetic fields in \KK\ theory
are described by five dimensional Minkowski space  with
twisted identifications. The four-dimensional reduction, however,
is only valid for four-dimensional
magnetic field parameter values $|B|<<1/R$ and for
distances from the axis of symmetry that are $< 1/|B|$. We justify the
latter by arguing that magnetic fields in the real world will have
finite
spatial extent.

We have demonstrated that the euclidean five-dimensional Kerr metric
gives an instanton describing the decay of
\KK\ magnetic fields, and argued that for a physical four-dimensional
magnetic field (i.e. $|B| << 1/R$) there are
two ways it can decay: by producing single singularities
into which space ``collapses,'' or by producing pairs of
monopole-anti-monopole pairs which accelerate off to infinity.
The former type of decay is much more likely, for small fields,
than the latter.
We have seen that this Kerr instanton is, in fact, identical to the
\KK\ Ernst instanton, and discussed several consequences of this surprising
fact.  We have also shown
that the fifth dimension tends to decompactify dynamically
after the second, rarer decay by pair production.

Thus we arrive at the final four-dimensional
 picture. If the magnetic field is zero, then
the vacuum decays by endlessly producing apparent
naked singularities. {}From
the five-dimensional point of view these correspond to Witten's
``bubbles of nothing'' which must eventually collide, and so in the
four dimensional description the singularities will coalesce. If we start at
time $t=0$ in the \KK\ vacuum (though it is hard to imagine, given
these instabilities, how we could have gotten into the vacuum in the
first place) then at any finite time, there is still an infinite
amount of flat space left and the process
will continue forever.  If there is any non-zero magnetic field present,
then, as well as this decay, there is
a small chance that a pair of monopoles will be pair created.

Since many currently popular unified theories include extra
spatial dimensions,
it is important to ask why the never-ending bubble nucleation
discussed here
is not a problem for these theories.
One resolution was proposed in \witten, and is applicable if
the theory contains fundamental fermions.
In five dimensional \KK\ theory, there
are inequivalent ways to include elementary
fermions; one
must specify the periodicity of the fermions around the compact direction
or in other words a spin structure.
When spacetime has topology $R^4\times S^1$
there are two choices for the spin structure (for simplicity,
we will assume that the fermions are not coupled to any extra $U(1)$ symmetry).
In the vacuum, the choices can be distinguished by asking how
spinors transform as they are parallelly transported around
the fifth direction. For one spin structure they come back
to themselves, for the other they pick up a minus sign.
As Witten pointed out, the five dimensional Schwarzschild
instanton has topology $R^2\times
S^3$ and consequently a unique spin structure. Asymptotically,
Schwarzschild tends to the \KK\ vacuum and one can ask which
spin structure one obtains there.
It turns out that a spinor picks up a minus sign under parallel
propagation around the fifth direction.
Thus, if one chooses the other spin structure for the vacuum (which is the
conventional choice - required to have
massless fermions and supersymmetry), it cannot decay via the
Schwarzschild instanton.

What about magnetic fields?  Since the Melvin solution again
has topology $R^4 \times S^1$, there are
two spin structures. Even though the spacetime is locally flat, the
nontrivial indentifications imply that if a vector is parallelly propagated
around the $S^1$, it will return rotated by an angle $2\pi RB$. It follows
that for one spin structure, parallel propagation of a spinor
around the fifth direction results in the spinor acquiring a
phase $e^{\pi R B \gamma}$, where $\gamma$ is a generator
of the Lie algebra of the spin group $\rm{Spin}(5)$ and
$\gamma^2 = -1$. For the
other   spin structure, parallel propagation gives a phase
$-e^{\pi R B \gamma}$.

Since
the topology of the Kerr instanton \kinst\ is again $R^2\times S^3$ it also
has a unique spin structure.
It tends to Melvin at infinity and one can show that
spinors  pick up the phase $-e^{\pi R {\alpha\over\mu} \gamma}$  under
parallel transport around the  closed
integral curves of
$l$, \kv, at infinity.
Now, for a given
four dimensional magnetic field of parameter $B$, there are two
instantons that describe its decay, as we discussed: \kinst\ with
(i) $\a/\mu = B$ and reduced along $l$
and (ii) $\a/\mu = B -1/R$
and reduced along
$\hat l = l + (1/R) \p_\vp$.
Instanton (i) has spinors which pick up the phase $-e^{\pi R B\gamma}$
when parallelly transported around orbits of $l$.
In the case of (ii) one might think that
the  extra
rotation involved in the definition of the internal direction
would introduce an extra minus sign into the phase.
This is not the case. Since the spacetime is almost flat  near
infinity, this extra rotation has the same effect as parallelly propagating a
spinor  around a circle in flat spacetime.
It does not introduce another minus sign.
Thus spinors on (ii)  pick up a phase  $-e^{\pi R (B - 1/R) \gamma}
= + e^{\pi R B \gamma}$
on parallel transport around orbits of $l'$ (and $l$). We see that the
spin structures on the two
different instantons correspond to the two inequivalent choices of
spin structure on the Melvin background. Thus choosing the
one in which spinors pick up the phase  $ e^{\pi R B \gamma}$
(which is the natural generalization, for small $B$, of the standard choice)
rules out decay via instanton (i), the ``bubble'' type decay,
but allows decay via (ii), the pair creation of monopoles.


It is natural to wonder what the implications of our results are for string
theory.
The Kaluza-Klein monopole solves the string equations of motion to leading
order in $\alpha'$. For large $R$ the five dimensional curvature is small
and we do not expect significant $\alpha'$ corrections \banks. Although it is
not yet known
how to determine the soliton spectrum in string theory, it is natural to assume
that the Kaluza-Klein monopole solution corresponds to a state in the Hilbert
space of toroidal compactifications (although the fact that this solution
does not approach the standard \KK\ vacuum at infinity \bomb\ is a subtlety
that needs to be addressed).
Supersymmetric toroidally compactified string theories are conjectured to
be invariant under strong weak coupling duality (see for example
\sen,\wit,\hullt).
The states dual to the Kaluza-Klein monopoles are electrically charged
string winding states. It would be
interesting to calculate the pair production rate for
the elementary string states in the first quantised theory and
compare this with the rate calculated using the space-time
instanton techniques employed here.
This is currently under investigation.

\bigskip
\bigskip
{\bf Acknowledgements}

We would like to thank  D. Kastor, D. Lowe, D. Marolf, R. Myers, A. Sen
A. Steif, A. Strominger, S. Surya and J. Traschen for helpful
conversations. We particularly thank R. Gregory, J. Harvey and R. Sorkin
for helping us to resolve the issue of spin structures.
JPG is supported in part by the U.S. Dept. of Energy
under Grant No. DE-FG03-92-ER40701. FD and GTH are supported in part by
NSF grant PHY-9008502.
This work was initiated at the Newton Institute for Mathematical Sciences and
completed at the Aspen
Center for Physics.

\listrefs
\end